\documentclass{A}

\draft

\begin{document}

\title{Modeling non local thermodynamic equilibrium plasma using the Flexible Atomic Code data }
\author{Bo \textsc{Han}\altaffilmark{1,2}, Feilu \textsc{Wang}%
\altaffilmark{1}, David \textsc{Salzmann}\altaffilmark{3}, Gang \textsc{Zhao}%
\altaffilmark{1}}
\altaffiltext{1}{Key laboratory of Optical Astronomy, National Astronomical
Observatories,\\ Chinese Academy of Sciences, Beijing 100012, China}
\altaffiltext{2}{University of Chinese Academy of Sciences, Beijing 100049,
China} \altaffiltext{3}{Weizmann Institute of Science, Rehovot, Israel}
\email{bhan@bao.ac.cn}
\maketitle

\begin{abstract}
We present a new code, RCF("Radiative-Collisional code based on FAC"),
which is used to simulate steady-state plasmas
under non local thermodynamic equilibrium condition, especially
photoinization dominated plasmas. RCF takes almost all of the radiative and
collisional atomic processes into rate equation to interpret the plasmas
systematically. The Flexible Atomic Code (FAC) supplies all the atomic data
 RCF needed, which insures calculating completeness and consistency of
atomic data. With four input parameters relating to the radiation source and
target plasma, RCF calculates the population of levels and charge states, as
well as potentially emission spectrum. In preliminary application, RCF successfully
reproduces the results of a photoionization experiment with reliable atomic data.
The effects of the most important atomic processes on the charge state
distribution are also discussed.
\end{abstract}

\SetRunningHead{Author(s) in page-head}{Running Head}

\section{Introduction}

Non local thermodynamic equilibrium (NLTE) exists in a wide variety of
astrophysical and laboratory plasmas. Examples of NLTE astronomical plasmas
are the stellar corona, interstellar nebulae and some other low density
ionized plasmas. X-ray satellites, such as Chandra and XMM-Newton, provided
large amount of high resolution spectra from astronomical objects, many of which are
 in NLTE. In laboratory, NLTE exists in laser produced plasmas,
tokamaks and Z-pinch based experiments.

In the present paper, we introduce a new NLTE plasma computer code, which we
called Radiative-Collisional code based on FAC (abbreviated RCF). RCF is a radiative-collisional code that includes
photoionization as well, thereby it is appropriate especially to
astrophysical plasmas.
 In the following we show its accuracy relative to other codes.

Several similar codes, which are being used for analysis of astrophysical
spectra, have already been published in the literature. Examples are GALAXY (%
\cite{1998JPhB...31.2129R}, \cite{2004PhRvL..93e5002F}, \cite%
{2006JQSRT..99..712F}, \cite{2004JPhB...37L.337R}), NIMP (\cite%
{1992JPhB...25.2745D}, \cite{2004JPhB...37L.337R}), FLYCHK (\cite%
{2003JQSRT..81..107C}), CLOUDY (\cite{1998PASP..110..761F}), XSTAR (\cite%
{1996ApJ...465..994K}, \cite{2004ApJS..155..675K}, \cite{2001ApJS..133..221K}%
, \cite{2001ApJS..134..139B}, \cite{2003ApJ...592..516B}), PhiCRE (\cite%
{2011ApJ...742...52S}, \cite{2011ApJ...742...53W}) and SASAL (\cite%
{2014ApJ...783..124L}). Some of these codes are used to interpret laboratory
experiments (GALAXY, FLYCHK, NIMP and PhiCRE), while others are used in
analysis of astrophysical spectra (CLOUDY and XSTAR), while RCF is designed
to be applicable to both of the conditions above.

The aim of this paper is to give a detail introduction of RCF and to present
its application to a phoionization experiment. Section 2 presents the model
and the method of to calculate the atomic rate coefficients. In section 3,
we apply RCF to reproduce the iron charge state distribution of a
photonionization experiment, together with a discussion of the importance of
the various atomic processes. A short summary is given in the last section.

\section{Rate equation and Atomic Data}

\subsection{Rate Equation}

RCF is a steady-state collisional-radiative optically-thin model. Its
rate-equation (\cite{Salzmann1998}) is
\begin{equation}
\frac{dN_{i,j}}{dt}=populating\;\,processes-depopulating\;processes=0
\end{equation}%
where $N_{i,j}$ is the density of the $j$th level of the $i$th charge state.
The processes included are the ionization and recombination between
neighboring charge states and excitation and de-excitation within the same
charge state. Their inverse processes are also taken into account by detailed
balance principle. The processes included in Eq.(1) are listed in Table \ref{tab:AP}.

\begin{longtable}{lll}
  \caption{Atomic processes in RCF (\cite{Salzmann1998}).}\label{tab:AP}
  \hline
  Reaction & Direct Process & Inverse Process  \\
\endfirsthead
  \hline
  Reaction & Direct Process & Inverse Process  \\
\endhead
  \hline
\endfoot
  \hline
\endlastfoot
  \hline
   $X_{i,j} \rightleftharpoons X_{i,j'} + h\nu$ & Spontaneous Decay & Photo excitation \\
   $X_{i,j} + e \rightleftharpoons X_{i,j'} + e'$ & Electron Impact Excitation & Electron Impact Deexcitation \\
   $X_{i,j} + h\nu \rightleftharpoons X_{i+1,j'} + e$ & Photoionization & Radiative Recombination \\
   $X_{i,j} + e \rightleftharpoons X_{i+1,j'} + e' + e''$ & Electron Impact Ionization & Three-Body Recombination \\
   $X_{i,j} \rightleftharpoons X_{i+1,j'} + e$ & Autoionization & Dieletronic Captrue
\end{longtable}

\subsection{Atomic Data and Reaction Rates}

The atomic data for RCF are calculated by FAC (\cite{2008CaJPh..86..675G}).
FAC is a fully relativistic software package that computes various atomic
data, which has been widely used in astrophysical and laboratory research (%
\cite{2008CaJPh..86..675G}). With the ions and their configurations as
input, the SFAC interface can supply $j-j$ coupled energy levels ($E_{i,j}$%
), bound-bound spontaneous decay rates ($A_{i,j\rightarrow j^{\prime }}$),
bound-bound electron-collision excitation (CE) cross sections ($\sigma _{CE}$%
), bound-free photoionization (PI) and electron-impact ionization (EI) cross
sections ($\sigma _{PI},\sigma _{EI}$), autoionization rate (AI) ($R_{AI}$),
and free-bound radiative recombination (RR) cross section ($\sigma _{RR}$),
where $\sigma _{PI}$ and $\sigma _{RR}$ are related through the Milne
relation.

The processes in Table 1 can be divided into two kinds, the inherent
reactions inside the plasma and the reactions driven by an external
radiation field.

The inherent ones include the reactions between ions and electrons driven by
collision, and spontaneous decay inside all ions. The solar corona plasma is
believed to be dominated by these processes. In such low density plasma, the
dominant processes are spontaneous decay and radiative recombination, whose
rates are much higher than collisional decay and three-body recombination. As a
result, the plasma departs from the local thermodynamic equilibrium, and can
no longer be described by Saha and Boltzmann equations. The collisions
between particles cause energy exchange and state distribution changes. Usually charge
exchange reactions between ions have negligibly low probabilities (\cite{Salzmann1998}).

The rate per volume of electron impact excitation reaction is calculated by
\begin{equation}
\frac{reactions}{cm^{3}s}=N_{i,j}n_{e}\mathcal{E}_{c},
\end{equation}%
where $n_{e}$ is the electron density in plasma, and $\mathcal{E}_{c}$(cm$%
^{3}$s$^{-1}$) is the collisional excitation rate coefficient. The rate
coefficient of collisional excitation is given by
\begin{equation}
\mathcal{E}_{c}=\int_{\Delta E}^{+\infty }f(v)\sigma _{CE}(v)dv,
\end{equation}%
where $f(v)$ is velocity distribution of electrons, assumed to have
Maxwellian distribution with electron temperature $T_{e}$, $\sigma _{CE}(v)$
is the collisional excitation cross section from $X_{i,j}$ to $%
X_{i,j^{\prime }}$ at velocity $v$, and $\Delta E$ is the excitation energy.
Thus, the CE rate coefficient expressed in term of incident electron energy
is
\begin{equation}
\mathcal{E}_{c}=\sqrt{\frac{8}{\pi m_{e}T_{e}^{3}}}\int_{\Delta E}^{+\infty
}E\sigma _{CE}(E)exp(-\frac{E}{T_{e}})dE,
\end{equation}%
where $E$ is incident electron energy, $m_{e}$ is the mass of electron, and $%
\sigma _{CE}(E)$ is the collisional excitation cross section calculated by
FAC.

The calculations of collisional ionization and radiative recombination rates
have a similar form with CE. The CI and RR rate coefficients (cm$^{3}$s$^{-1}
$) are
\begin{equation}
\mathcal{I}_{c}=\sqrt{\frac{8}{\pi m_{e}T_{e}^{3}}}\int_{\Delta E}^{+\infty
}E\sigma _{CI}(E)exp(-\frac{E}{T_{e}})dE,
\end{equation}%
\begin{equation}
\mathcal{R}_{r}=\sqrt{\frac{8}{\pi m_{e}T_{e}^{3}}}\int_{0}^{+\infty
}E\sigma _{RR}(E)exp(-\frac{E}{T_{e}})dE,
\end{equation}%
where $\sigma _{CI}(E)$ is the CI cross section from $X_{i,j}$ to $%
X_{i+1,j^{\prime }}$ and $\sigma _{RR}(E)$ is the RR cross section from $%
X_{i+1,j^{\prime \prime }}$ to $X_{i,j}$.

The inverse process to CI is three-body recombination from $X_{i+1,j^{\prime
\prime }}$ to $X_{i,j}$ and CE's inverse process is collisional deexcitation
from $X_{i,j^{\prime }}$ to $X_{i,j}$. Their rate coefficients are obtained
by Detailed Balance Principle
\begin{equation}
\mathcal{R}_{t}=\frac{1}{2}\left( \frac{2\pi ~\hbar ^{2}}{m_{e}T_{e}}\right)
^{3/2}\frac{g_{i,j}}{g_{i+1,j^{\prime }}}exp(\frac{\Delta E}{T_{e}})\mathcal{%
I}_{c},
\end{equation}%
\begin{equation}
\mathcal{D}_{c}=\frac{g_{i,j}}{g_{i,j^{\prime \prime }}}exp(\frac{\Delta E}{%
T_{e}})\mathcal{E}_{c},
\end{equation}%
It should be mentioned that the three body recombination rate coefficient,
being dependent on two electrons, is proportional to $n_{e}^{2}$,
\begin{equation}
\frac{reactions}{cm^{3}s}=N_{i,j}{n_{e}}^{2}\mathcal{R}_{t},
\end{equation}%
and the unit of $\mathcal{R}_{t}$ is cm$^{6}$s$^{-1}$.

The spontaneous processes in plasma are radiative decay and autoionization.
Their rates are directly given by FAC, $A_{i,j\rightarrow j^{\prime }}$(s$%
^{-1}$) and $R_{AI}$(s$^{-1}$). Radiative decay process is the mechanism of
plasma emitting line spectrum. In the present work, we assume that
the photons emitted by radiative decay are not reabsorbed.
 Autoionization occurs in case of doubly excited ions, and is
possible only if the sum of the energies of the two electrons is higher than
the binding energy of the ion. During autoionization, the energy, which is released by the inner
excited electrons decays to a lower state, ionizes the outer one into the
continuum. There are several ways to produce such highly excited ions, such
as dielectron capture, photoexcitation and photoionization of inner-shell
electrons, and the two inner-shell processes are especially important in
conditions where strong fields exist. Dielectronic capture into doubly
excited states is the inverse process of autoionization, and its rate
coefficient (cm$^{3}$s$^{-1}$) is obtained by the detailed balance
principle,
\begin{equation}
\mathcal{R}_{d}=\frac{1}{2}\left( \frac{2\pi ~\hbar ^{2}}{m_{e}T_{e}}\right)
^{3/2}\frac{g_{i,j}}{g_{i+1,j^{\prime }}}exp(\frac{\Delta E}{T_{e}})R_{AI}.
\end{equation}%
The rate coefficient of dielectronic recombination is obtained when Eq.(10)
is multiplied by the branching ratio for radiative stabilization of the
doubly excited state, $A_{i,j}/(\sum A_{i,j}+\sum R_{AI})$ (\cite%
{Salzmann1998}).

When the plasma is irradiated by a strong external radiation source, the
radiative field will excite or ionize the ions causing the plasma gets into
photoionizational collisional radiative equilibrium regime. In this case,
the photoionization and photoexitation processes are not negligible and may
dominate the charge state distribution. For example, the strong radiation
field emitted by an accreting compact object is believed to be the main
ionization mechanism of the highly ionized low density gas around it.

The photoionization reaction rate per unit volume is calculated by
\[
\frac{reactions}{cm^{3}s}=N_{i,j}R_{PI}=N_{i,j}\int_{\Delta E}^{+\infty
}n_{p}(h\nu )c\sigma _{PI}(h\nu )d(h\nu ),
\]%
where $R_{PI}$ is the photoionization rate(s$^{-1}$), $\Delta E$ is the
ionizing energy from $X_{i,j}$ to $X_{i+1,j^{\prime }}$, $h\nu $ is the
energy of incident photon, $\sigma _{PI}(h\nu )$ is the photoionization
cross section, c is the speed of light, and $n_{p}(h\nu )\,d(h\nu )$ is the
density of photons having energy in the range $[h\nu ,h\nu +d(h\nu )]$. For a
black body radiation source having radiation temperature $T_{r}$, with
energy intensity $I(h\nu )$ (eV/(cm$^{2}\cdot $s$\cdot $ eV)) and dilution
factor $\alpha $, $n_{p}(h\nu )=(\alpha I(h\nu ))/(ch\nu )$, then $%
R_{PI}$ becomes
\begin{equation}
R_{PI}=\alpha \int_{\Delta E}^{+\infty }\frac{I(h\nu )}{h\nu }\sigma
_{PI}(h\nu )d(h\nu ),
\end{equation}

The photoexcitation rate per volume from $X_{i,j}$ to $X_{i,j^{\prime }}$ is
\begin{equation}
\frac{reactions}{cm^{3}s}=N_{i,j}R_{PE}=N_{i,j}hJ(h\nu )B_{i,j\rightarrow
j^{\prime }},
\end{equation}%
$B_{i,j\rightarrow j^{\prime }}$ is the Einstein $B$-coefficient The
photo-excitation rate($s^{-1}$) irradiated by a diluted isotropical black
body radiation source is
\begin{equation}
R_{PE}=\alpha ~I(h\nu )\frac{g_{j^{\prime }}}{g_{j}}\frac{h^{2}c^{2}}{2(h\nu
)^{3}}A_{i,j^{\prime }\rightarrow j},
\end{equation}

Currently, RCF is assumes a blackbody radiator, thereby the photoionization
and photoexcitation rate are
\begin{equation}
R_{PI} =\alpha \frac{2}{h^{3}c^{2}}\int_{\Delta E}^{+\infty }\frac{(h\nu)^{2}%
}{e^{\frac{h\nu}{T_{r}}}-1}d(h\nu) ,
\end{equation}
\begin{equation}
R_{PE}=\alpha ~\frac{g_{j^{\prime}}}{g_{j}}A_{i,j^{\prime}\rightarrow j}%
\frac{1}{e^{\frac{h\nu}{T_{r}}}-1} ,
\end{equation}

\subsection{Input Parameters}

These equations require four input parameters in RCF, which are radiation
temperature $T_{r}$, dilution factor $\alpha $, electron temperature $T_{e}$%
, and electron density $n_{e}$. $T_{r}$ is the temperature of the blackbody
radiation source. $\alpha $ stands for the attenuation of radiation between
the source and the irradiated plasma, and it mainly depends on the opacity
and distance. $T_{e}$ and $n_{e}$ are the properties of the irradiated
plasma, and they are sufficient to describe plasma under coronal
equilibrium. However, when the external radiation field is important, all
the four input parameters are needed.

In experiments these parameters are measured directly or deduced indirectly
from some measured values. However, depending on the experiment setup, some
parameters cannot be obtained. For example, the  experiment by Fujioka et
al. (2009) with silicon target provided all four parameters with some
uncertainties, whereas $T_{e}$ does not have a definite value in the Sandia
experiment by Foord et al. (2004) with iron target. In astrophysics, $T_{r}$
is estimated by the observed continuum spectrum, and $\alpha $ is roughly
deduced from the distance between two celestial objects by the inverse
square law. Usually, $T_{e}$ and $n_{e}$ of the irradiated plasma are
deduced from some characteristic spectral line ratios. For example, the
ratios of resonance, intercombination and forbidden lines of He-like ions
are important diagnostics of electron density and temperature (\cite%
{2000A&AS..143..495P}). The radiative recombination continuum (RRC) is also
an important method to diagnose electron temperature of plasmas.

\section{Simulation of Sandia Photoionization Experiment}

Photoionized plasma is a special kind of NLTE plasma. It is widely observed
in the universe, such as low density nebula near accreting X-ray source.
Recently, this kind of plasmas were also produced in laboratory using high
power laser (\cite{2009NatPh...5..821F}) and Z-pinch (\cite%
{2004PhRvL..93e5002F}).

In this section, RCF is applied to the photoionization experiment at Sandia
National Laboratory Z-facility (\cite{2004PhRvL..93e5002F}). In this
experiment, a 165 eV near-blackbody radiation source was created to produce
a $n_{e}=2\pm 0.7\times 10^{19}$cm$^{-3}$ plasma (\cite{2004PhRvL..93e5002F}%
) in photoionizational collisional radiative equilibrium regime (\cite%
{2011ApJ...742...53W}). A distribution of iron charge states was deduced
from the absorption spectrum. A number of papers (\cite{2004PhRvL..93e5002F}%
, \cite{2006JQSRT..99..712F}, \cite{2011ApJ...742...53W}, \cite%
{2013JPSJ...82b4501H}, \cite{2014ApJ...783..124L}) tried reproducing the
measured charge state distribution using different models and computer
codes. All these works assumed a steady state photoionized plasma, which is
also adopted by RCF.

In this experiment, only two of the parameters needed in RCF are specified,
which are $T_{r}$ and $n_{e}$. $T_{e}$ is a disputed focus of the former
works, and it spans from 70 eV to 150 eV in different models (\cite%
{2006JQSRT..99..712F}, \cite{2011ApJ...742...53W}). However, $\alpha $ was
not specified by some models (\cite{2006JQSRT..99..712F}), although it is an
important parameter that controls the influence of radiative field on the
plasma (\cite{2013JPSJ...82b4501H}). Fortunately, the experiment yielded an ionization parameter $\xi
_{exp}=25$ $erg\cdot cm\cdot s^{-1}$ (\cite{2004PhRvL..93e5002F}) at the
peak of the radiation pulse. $\xi $ is a parameter related to the radiation
field $\xi =16\pi ^{2}J/n_{e}$, and for an isotropic blackbody radiation
field $\xi =16\pi \sigma T_{r}^{4}/n_{e}$, where $J$ is the mean intensity
and $\sigma $ is the Stefan-Boltzmann constant. Therefore, the ratio between
experimental value and theoretical value stands for the attenuation of the
radiative field, ie. $\alpha =\xi _{exp}/\xi _{theo}$, which can be derived
as $\alpha =0.85\%-1.76\%$.

\begin{figure}[tbp]
\begin{center}
\includegraphics[width=\textwidth]{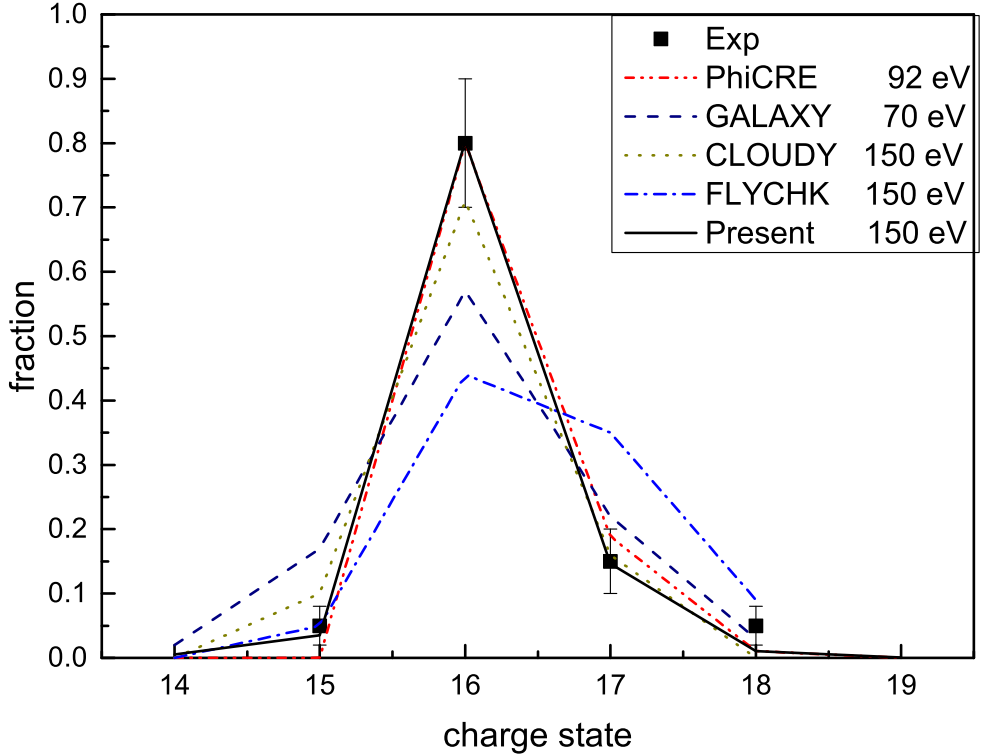}\\[0pt]
\end{center}
\caption{The charge state distribution calculated by RCF (Solid) for Fe
photoionization experiment(Scatters), and comparison with PhiCRE (Dash Dot
Dot), GALAXY (Dash), CLOUDY (Dot), and FLYCHK (Dash Dot) (\protect\cite%
{2006JQSRT..99..712F}, \protect\cite{2011ApJ...742...53W}). The temperatures
are electron temperatures used by these models. }
\label{f1}
\end{figure}

Figure \ref{f1} displays the charge state distribution of the iron photoionization
experiment predicted by RCF and comparisons with the experiment values and
some previous works (\cite{2006JQSRT..99..712F}, \cite{2011ApJ...742...53W}%
). Seemingly, RCF produces the result closest to the experiment, and almost
every ion is within the experiment uncertainties. The average charge state
in present calculation is $<Z>=16.12$, which agrees well with the measured
value $<Z>=16.1\pm 0.2$. The input parameters used here are $T_{r}$=165 eV, $%
n_{e}=2\times 10^{19}$cm$^{-3}$, $\alpha =1.4\%$ and $T_{e}=150eV$. $T_{e}$%
=150 eV agrees with CLOUDY and FLYCHK, and some other works, such as NIMP (%
\cite{2004JPhB...37L.337R}) and Han et al. (2013). $\alpha =1.4\%$ is in the
interval deduced above.

A main reason of the differences among the codes in Figure \ref{f1} is the
different sources of atomic data. GALAXY employs an average-of-configuration
approximation for electronic states, screened hydrogenic for both
collisional and radiative processes, and Hartree-Dirac-Slater or Kramers
cross sections for photoionization (\cite{1998JPhB...31.2129R}, \cite%
{2006JQSRT..99..712F}). FLYCHK uses hydrogenic approximation to calculate
energy levels and level populations (\cite{2003JQSRT..81..107C}, \cite%
{2006JQSRT..99..712F}). Results of GALAXY and FLYCHK largely deviate from
the measured one. The atomic databases of CLOUDY are accurate enough to be
comparable with spectral emission data (\cite{1998PASP..110..761F}, \cite%
{2006JQSRT..99..712F}), but there still are some obvious disparities between
it and the experiment. The energy levels and spontaneous decay rates of
PhiCRE are taken from the NIST database, and other rate coefficients are
calculated by widely used formulas (\cite{2011ApJ...742...52S}, \cite%
{2011ApJ...742...53W}).

\begin{figure}[tbp]
\begin{center}
\includegraphics[width=\textwidth]{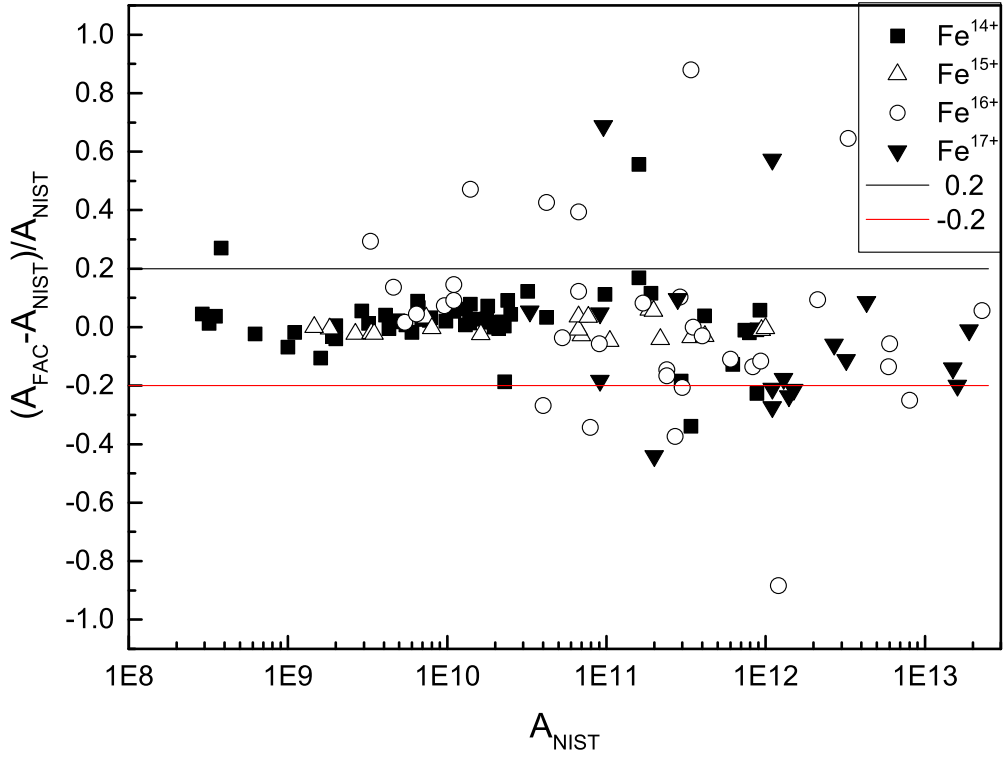}\\[0pt]
\end{center}
\caption{Comparison of radiative decay rates between present work and corresponding available data on NIST from Fe$^{14+}$ to Fe$^{17+}$.}
\label{f4}
\end{figure}

\begin{figure}[tbp]
\begin{center}
\includegraphics[width=\textwidth]{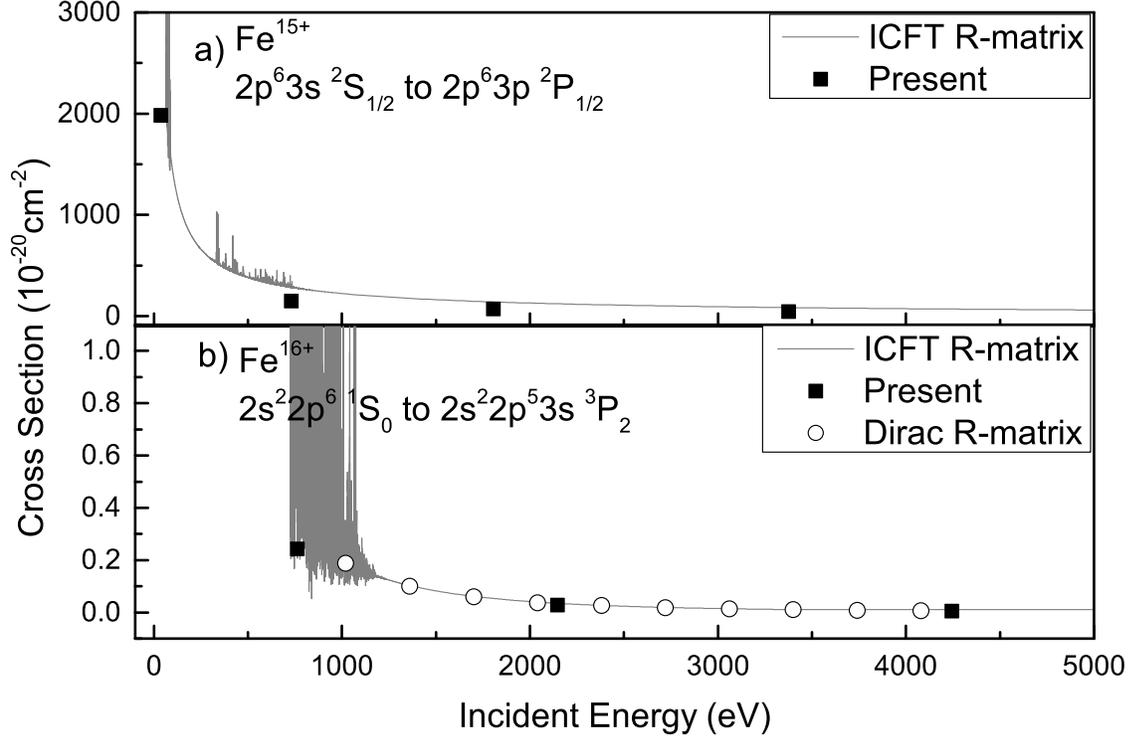}\\[0pt]
\end{center}
\caption{Comparison of collisional excitation cross sections. (a)Transition from $2p^{6}3s$ $^{2}S_{1/2}$ to
$2p^{6}3p$ $^2P_{1/2}$ of Fe$^{15+}$. Grey solid: ICFT R-matrix (\cite{2009A&A...500.1263L}), black square: present FAC.
 (b)Transition from $2s^{2}2p^{6}$ $^{1}S_0$ to $2s^{2}2p^{5}3s$ $^3P_2$ of Fe$^{16+}$. Grey solid: (\cite{2010A&A...518A..64L}),
  black square: present FAC, circle: Dirac R-matrix (\cite{2003ApJS..144..169A}).}
\label{f5}
\end{figure}

The atomic data of RCF are calculated by FAC, which calculates all the atomic data
by a fully relativistic approach based on Dirac equation (\cite%
{2008CaJPh..86..675G}), and this single theoretical framework ensures
self-consistency between different parts.
The configurations calculated with FAC of present work
 are listed Table \ref{tab:CON}, which include 1948 singly or doubly excited levels.
 For saving computation time, the maximum principle quantum number n is set to be 4.
Because the states with K-shell vacancies have energies higher than 7 keV, which are much higher than
the energies of photons and electrons under this experiment condition, K-shell is closed in FAC
calculation.
For ensuring the accuracy of present work, we compare the present FAC data with some literature values.
Table \ref{tab:lev} is the comparison of energy levels for $2s^22p^53l$ and $2s2p^63l$
states of Fe$^{16+}$ between the NIST database (\cite{NIST}) and present work, and it shows excellent agreement (within 0.4\%).
Figure \ref{f4} shows the comparison of radiative decay rates (s$^{-1}$) between present work and the available
corresponding transitions on the NIST database (\cite{NIST}) of the four most abundant charge states.
As shown, there are more than 80\% of present data within 20\% agreement with NIST database (\cite{NIST}).
According to Eq.(14), the accuracy of radiative decay rates also guarantees the calculating of photoexcitation rates.
For collisional excitation, Figure \ref{f5} is the comparison of cross section of Fe$^{15+}$ and Fe$^{16+}$
for transitions from their ground states to their first excited levels.
The present FAC results agree well with those calculated with ICFT (intermediate-coupling frame transformation) R-matrix (\cite{2009A&A...500.1263L},\cite{2010A&A...518A..64L})
 and Dirac R-matrix (\cite{2003ApJS..144..169A}).
 Using these data, RCF successfully reproduces the experiment result, and we look forward to
 applying it to spectral analysis of laboratory or astrophysical plasmas in future works.

\begin{longtable}{llllll}
  \caption{The configurations used by Case A.}\label{tab:CON}
  \hline
  Charge State & Singly Excited & ~ & ~ &  Doubly Excited & ~  \\
\endfirsthead
  \hline
  Charge State & Singly Excited & ~ & ~ &  Doubly Excited & ~  \\
\endhead
  \hline
\endfoot
  \hline
\endlastfoot
  \hline
   $Fe^{14+}$ & $2s^{2}2p^{6}3s^{2}$ & $2s^{2}2p^{6}3s3p$ & $2s^{2}2p^{6}3p^{2}$ &  $2s^{2}2p^{5}3s^{2}3p$ & $2s2p^{6}3s^{2}3p$ \\
        ~     & $2s^{2}2p^{6}3s3d$   & $2s^{2}2p^{6}3p3d$ & $2s^{2}2p^{6}3d^{2}$ &           ~             &        ~           \\
        ~     & $2s^{2}2p^{6}3s4s$   & $2s^{2}2p^{6}3s4p$ & $2s^{2}2p^{6}3s4d$   &           ~             &        ~           \\
        ~     & $2s^{2}2p^{6}3s4f$   &          ~         &         ~            &           ~             &        ~           \\
   $Fe^{15+}$ & $2s^{2}2p^{6}3s$     & $2s^{2}2p^{6}3p$   & $2s^{2}2p^{6}3d$     & $2s^{2}2p^{5}3s^{2}$ & $2s^{2}2p^{5}3s3p$   \\
        ~     & $2s^{2}2p^{6}4s$     & $2s^{2}2p^{6}4p$   & $2s^{2}2p^{6}4d$     & $2s^{2}2p^{5}3p^{2}$ & $2s^{2}2p^{5}3s3d$   \\
        ~     & $2s^{2}2p^{6}4f$     &          ~         &         ~            & $2s^{2}2p^{5}3p3d$   & $2s^{2}2p^{5}3d^{2}$ \\
        ~     &           ~          &          ~         &         ~            & $2s2p^{6}3s^{2}$     & $2s2p^{6}3s3p$       \\
        ~     &           ~          &          ~         &         ~            & $2s2p^{6}3p^{2}$     & $2s2p^{6}3s3d$       \\
        ~     &           ~          &          ~         &         ~            & $2s2p^{6}3p3d$       & $2s2p^{6}3d^{2}$     \\
   $Fe^{16+}$ & $2s^{2}2p^{6}$       & $2s^{2}2p^{5}3s$   & $2s^{2}2p^{5}3p$     & $2s2p^{5}3s^{2}$     & $2s2p^{5}3s3p$       \\
        ~     & $2s^{2}2p^{5}3d$     & $2s2p^{6}3s$       & $2s2p^{6}3p$         & $2s2p^{5}3p^{2}$     & $2s2p^{5}3s3d$       \\
        ~     & $2s2p^{6}3d$         & $2s^{2}2p^{5}4s$   & $2s^{2}2p^{5}4p$     & $2s2p^{5}3p3d$       & $2s2p^{5}3d^{2}$     \\
        ~     & $2s^{2}2p^{5}4d$     & $2s^{2}2p^{5}4f$   &         ~            &          ~           &           ~          \\
   $Fe^{17+}$ & $2s^{2}2p^{5}$       & $2s2p^{6}$         & $2s^{2}2^{4}3s$      & $2s2p^{4}3s^{2}$     & $2s2p^{4}3s3p$       \\
        ~     & $2s^{2}2^{4}3p$      & $2s^{2}2^{4}3d$    & $2s2p^{5}3s$         & $2s2p^{4}3p^{2}$     & $2s2p^{4}3s3d$       \\
        ~     & $2s2p^{5}3p$         & $2s2p^{5}3d$       & $2s^{2}2^{4}4s$      & $2s2p^{4}3p3d$       & $2s2p^{4}3d^{2}$     \\
        ~     & $2s^{2}2^{4}4p$      & $2s^{2}2^{4}4d$    & $2s^{2}2^{4}4f$      &          ~           &           ~          \\
        ~     & $2s^{2}2^{4}4s$      & $2s^{2}2^{4}4p$    & $2s^{2}2^{4}4d$      &          ~           &           ~          \\
        ~     & $2s^{2}2^{4}4f$      &         ~          &         ~            &          ~           &           ~          \\
   $Fe^{18+}$ & $2s^{2}2p^{4}$       & $2s2p^{5}$         & $2p^{6}$             &          ~           &           ~          \\
        ~     & $2s^{2}2p^{3}3s$     & $2s2p^{4}3s$       &         ~            &          ~           &           ~          \\
   $Fe^{19+}$ & $2s^{2}3p^{3}$       &         ~          &         ~            &          ~           &           ~
\end{longtable}

\begin{longtable}{lllll}
  \caption{Comparison of energy levels for $2s^22p^53l$ and $2s2p^63l$ states of Fe$^{16+}$ between NIST and present work.}\label{tab:lev}
  \hline
  Index & Level & ~  & NIST & Present   \\
\endfirsthead
  \hline
  Index & Level & ~ & NIST & Present   \\
\endhead
  \hline
\endfoot
  \hline
\endlastfoot
  \hline
 0 & $2s^{2}2p^{6}$	  & $^{1}S_{0}$   &		0	    &	0     \\
 1 & $2s^{2}2p^{5}3s$ & $^{3}P_{2}$   & 725.2443	&	723.9074\\
 2 &        ~         & $^{1}P_{1}$   & 727.1388	&	725.874\\
 3 & $2s^{2}2p^{5}3s$ & $^{3}P_{0}$   & 737.856	    &	736.5231\\
 4 &         ~        & $^{3}P_{1}$   & 739.0537	&	737.762\\
 5 & $2s^{2}2p^{5}3p$ & $^{3}S_{1}$   & 755.4915	&	755.008\\
 6 & $2s^{2}2p^{5}3p$ & $^{3}D_{2}$   & 758.9928	&	757.7623\\
 7 &                  & $^{3}D_{3}$   & 760.6095	&	759.327 \\
 8 &                  & $^{3}D_{1}$   & 771.0614	&	769.7896\\
 9 & $2s^{2}2p^{5}3p$ & $^{1}P_{1}$   & 761.7403	&	760.5093\\
10 & $2s^{2}2p^{5}3p$ & $^{3}P_{2}$   & 763.5529	&	762.2596\\
11 &                  & $^{3}P_{0}$   & 768.981	    &	767.8495\\
12 &                  & $^{3}P_{1}$   & 774.3073	&	773.1159\\
13 & $2s^{2}2p^{5}3p$ & $^{1}D_{2}$   & 774.6855	&	773.415\\
14 & $2s^{2}2p^{5}3p$ & $^{1}S_{0}$   & 787.7224	&	790.3189\\
15 & $2s^{2}2p^{5}3d$ & $^{3}P_{0}$   & 801.4313	& 	800.4122\\
16 &                  & $^{3}P_{1}$   & 802.401	    &	801.3396\\
17 &                  & $^{3}P_{2}$   & 804.211	    &	803.0749\\
18 & $2s^{2}2p^{5}3d$ & $^{3}F_{4}$   & 804.2644	&	802.9189\\
19 &                  & $^{3}F_{3}$   & 805.0331	&	803.6367\\
20 &                  & $^{3}F_{2}$   & 817.5964	&	816.2671\\
21 & $2s^{2}2p^{5}3d$ & $^{1}D_{2}$   & 806.728	    &	805.3275\\
22 & $2s^{2}2p^{5}3d$ & $^{3}D_{3}$   & 807.8004	&	806.4032\\
23 &                  & $^{3}D_{1}$   & 812.369	    &	811.2368\\
24 &                  & $^{3}D_{2}$   & 818.4135	&	817.0462\\
25 & $2s^{2}2p^{5}3d$ & $^{1}F_{3}$   & 818.9342	&	817.4908\\
26 & $2s^{2}2p^{5}3d$ & $^{1}P_{1}$   & 825.7		&   825.4368\\
27 & $2s2p^{6}3s$     & $^{1}S_{0}$   & 869.1		&   867.1547\\
28 & $2s2p^{6}3p$     & $^{3}P_{1}$   & 892.55	    &	892.5898\\
29 & $2s2p^{6}3p$     & $^{1}P_{1}$   & 896.939	    &	895.3807
\end{longtable}

\begin{figure}[tbp]
\begin{center}
\includegraphics[width=\textwidth]{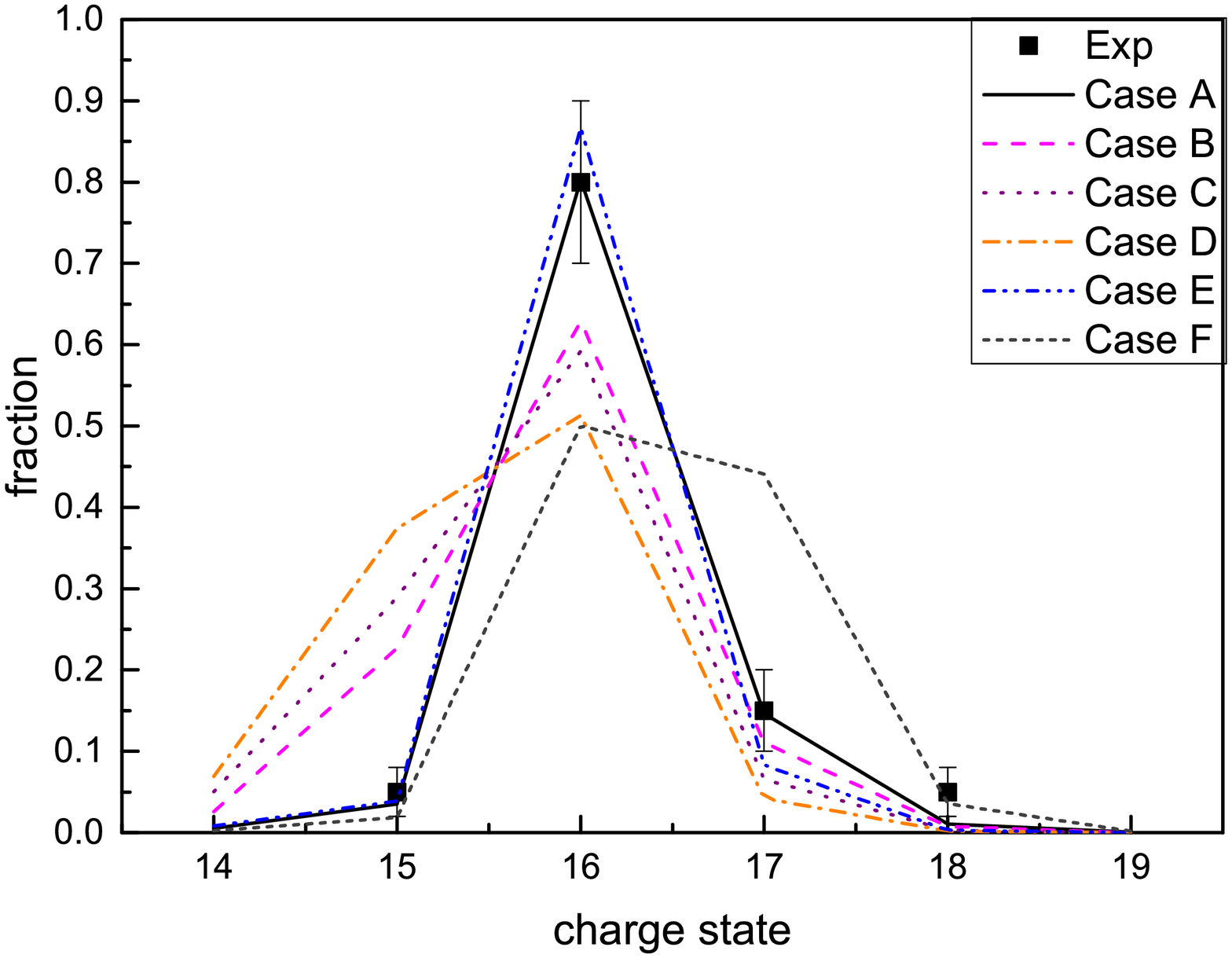}\\[0pt]
\end{center}
\caption{The charge state distribution calculated by RCF with different data
or without specific processes. Case A uses data with inner-shell holes
(Solid), Case B use data without inner-shell holes (Dash).Case C (Dot), D
(Dash Dot), E (Dash Dot Dot) use the same data as Case A, but there are no
AI and DC processes in Case C, no PE process in Case D, no PI process in
Case E, no CE process in Case F. The scatters are experiment values. }
\label{f2}
\end{figure}

In Figure \ref{f2}, we compare the influences of some important processes on
the experimental results. Case A is the one presented in Figure \ref{f1}, in which
we included all the relevant processes in RCF and all the configurations in Table \ref{tab:CON}.

Case B uses data without the doubly excited levels in Table \ref{tab:CON}. As can be seen, there are
significant differences relative to Case A.

In Case C both the autoionization and its inverse process, dielectronic
capture, are turned off. Results of Case C are close to those of Case B, and
both predict an average charge lower than that of Case A. This means that
doubly excited states are non-ignorable in the current calculation of charge state
distributions. In other words, autoionization is an important ionizing
channel, and doubly excited states act like ladders to the next charge state in this experiment.

\begin{figure}[tbp]
\begin{center}
\includegraphics[width=\textwidth]{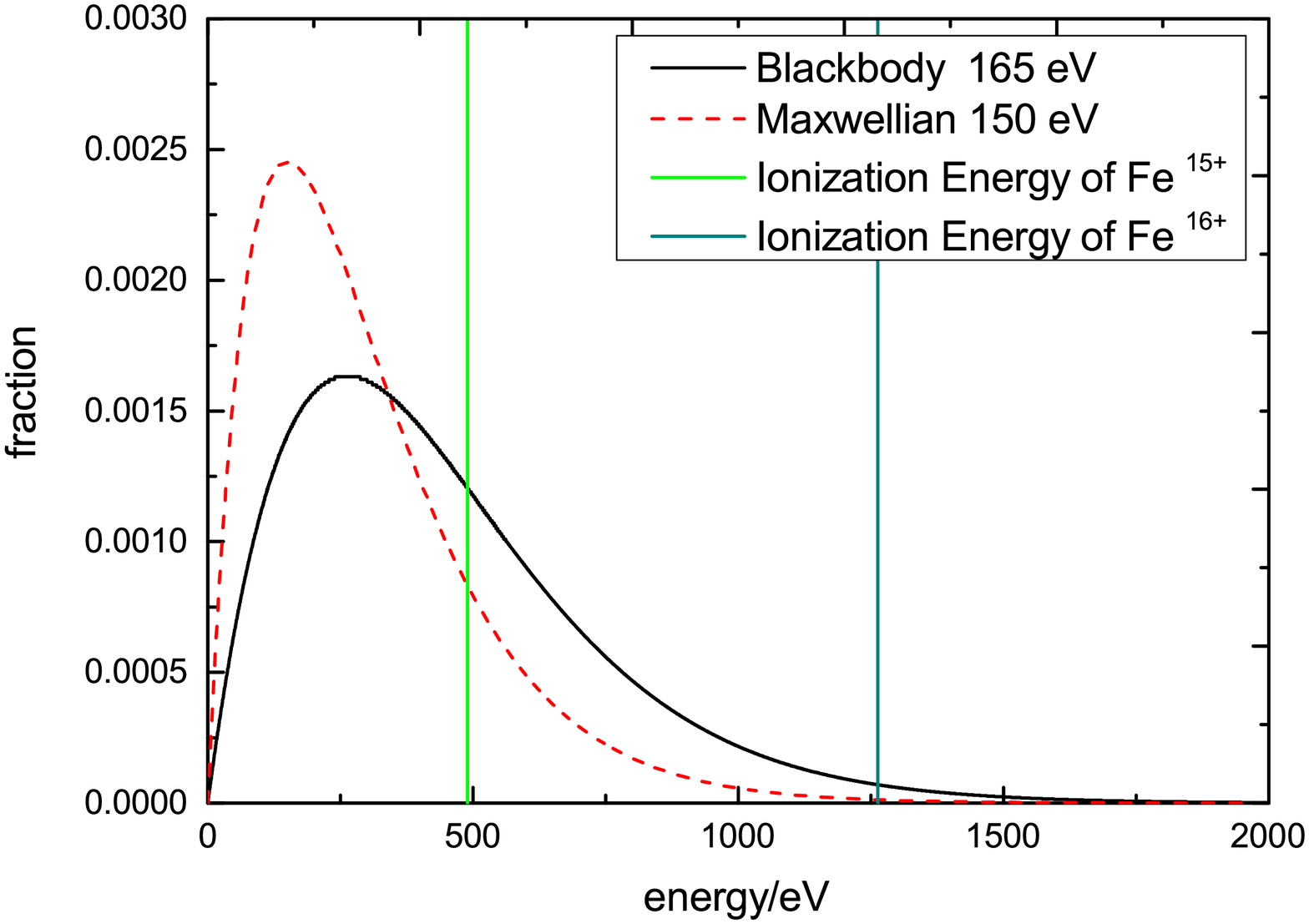}\\[0pt]
\end{center}
\caption{The distribution of 165 eV blackbody photons (solid) and 150 eV
Maxwellian electrons (dash). }
\label{f3}
\end{figure}

In Cases D and E, we examine the influence of the radiation field on the
Sandia experiment charge state distribution. First we note that there is a
big difference between the binding energy of the two most important charge
states in the experiment: the ionization energy of  Fe$^{15+}$ (Na-like) is
489.312 eV,  and that of  Fe$^{16+}$(Ne-like) is 1262.7 eV. As shown in
Figure \ref{f3}, a larger fraction of photons than electrons has sufficient energy to
excite or ionize Fe$^{+15}$ and Fe$^{+16}$. In fact, we have found that the
influence of electron collisional ionization is so small that its omission
from the computations is hardly different from Case A.

There are two radiation driven processes in the code, photoexcitation (PE) and
photoionization (PI). PE is omitted in Case D, whereas PI is turned off in
Case E. It can be seen from Figure \ref{f2} that the absence of PE significantly
reduces the average charge state, \textit{i.e.}, photoexcited states play an
important role in the charge state distribution. Case E is very close to
Case A. The reason for this behavior is the large difference between the
binding energies of Fe$^{+15}$ and Fe$^{+16}$. On the other hand, there is a
threshold between Fe$^{16+}$ and Fe$^{17+}$, because neon-like Fe$^{+16}$ has
a closed shell stable configuration.

Both of PE and PI have preference to ionize inner-shell electrons resulting
in autoionizing states. However, according to the photon energy distribution
in Figure \ref{f3} and the energy levels of main ions, the doubly excited levels
 seem to be more likely produced by PE process. Actually, when the PE channel
 to doubly excited states is shut down, the result is almost the same as Case D,
 which confirms that PE is the main pumping mechanism of doubly excited states.
 Case E indicates that PE+AI process wins the
competition in ionizing Fe$^{15+}$, but for Fe$^{16+}$ the PI channel is
important, too.

In Case F, collisional excitation (CE) is omitted, and it gets a strange
result. The deletion of CE does not lower $<Z>$ as Case D, but rather the
plasma is more ionized than Case A.
According to Figure \ref{f3}, although the electrons have comparatively lower energy
and cannot ionize the ions as effectively as photons,
they still can excite the ions to singly excited levels by collision.
However, the collision produced singly excited levels have smaller reaction
cross sections with photons than the ground state and lower levels, namely, they
are more difficult to be ionized by photons.
Therefore, the electrons certainly would take part in the competition of reacting with the
ground state of ions, and as a result reducing the ionizing efficiency of PI and PE+AI.
What is more, when the CE channel from the ground states of ions to
the singly excited states is shut down,
it produces a result similar to Case F, which confirms the discussion above.
So that, it makes sense that why $<Z>$ rises when CE is shut down.

In conclusion, RCF has a good agreement to the photoionization
experiment results (\cite{2004PhRvL..93e5002F}), and gives reasonable explanation for the charge state distribution.
 In the calculations of RCF, the charge state
distribution of this experiment is a composite result of different atomic
processes. The external field dominates the ionizations in the plasma by
photoionization directly and photoexcitation plus autoionization indirectly.
The transitions within any given single charge state can significantly
affect the charge state distribution, and one of the interesting results of
our computations is the role played by collisional excitation in this
experiment, in which it reduces the total ionization rate by competing with
PE and PI.

\section{Summary}

In this paper we introduced a new code, RCF, which is applied to plasma in
NLTE condition, especially in the photoionization dominating regime. This
code can calculate the level population, charge state distribution and
spectra of a plasma in steady state. The atomic data source of this code is
FAC, which is an easy to use and powerful software package to calculate
various atomic data. The SFAC interface can provide all the atomic data
needed in RCF, without any additional modifications. FAC is based on a fully
relativistic theoretical framework, which ensures the accuracy and
consistency of the atomic data.

All the plasma processes and their inverse ones are related by the detailed
balance principle in RCF. As a result, in high density regime, the RCF
generates results similar to the Saha equation with same atomic data. In
other words, RCF converges to LTE approximation under the appropriate
condition. In radiation dominant regimes, RCF gets a charge state
distribution which closely agrees to the results of the Fe photoionization
experiment. A comparison is given to the results of other similar codes. We
also discussed the influence of the various atomic processes to the charge
state distribution of this experiment. Photoionization is not the only
important ionizing channel, but the photoexcitation plus autoionization
process are proved to be also significant. Although the electrons
have comparatively lower energy than the photons, they still are important.
 The electrons can excite the ions to the levels which have small
reaction cross sections for the photons, and the result is reducing the
ionizing efficiency of the photons.

The charge state and levels' distributions are a prerequisite for the
simulation of the emission spectrum, and we showed that all the atomic
processes may have significant effect under the appropriate conditions. In
particular, in the analysis of X-ray spectrum from a compact object
photoexcitation is an important pumping mechanism (\cite{2014ApJ...780..121K}%
). Porquet \& Dubau (2004) also emphasized the influence of cascading decay
from higher levels and collisional excitation on the line ratios in plasma
diagnosis. According to the result shown, RCF is a reasonable code to get
accurate distributions in steady NLTE plasma by including all the processes
and using FAC data. We shall use it for spectrum analysis in astrophysics
 and laboratory of photoionizing and collisional NLTE plasmas in our further
works.


\bigskip

Acknowledgement

This work is supported by the NSFC under grants Nos.11173032 and 11135012,
and by the National Basic Research Program of China (973 Program) under
grant No.2013CBA01503. 

\end{document}